\documentclass[twocolumn,showpacs,preprintnumbers,amsmath,amssymb,aps]{revtex4}

\usepackage{graphicx}
\usepackage{dcolumn}
\usepackage{bm}

\begin{document}


\author{Peter B. Weichman}

\affiliation{$^2$ALPHATECH, Inc., 6 New England Executive Place,
Burlington, MA 01803}

\title{Universal early-time response in high-contrast
electromagnetic scattering}

\date{\today}

\begin{abstract}

The time-domain response of highly conducting targets following a
rapidly terminated electromagnetic pulse displays three distinct
regimes: early, intermediate and late time.  The intermediate and
late times are characterized by a superposition of exponentially
decaying eigenmodes.  At early time an ever increasing number of
rapidly decaying modes contribute, with the result that the
scattered electric field displays a universal $t^{-1/2}$ power law
which emerges from the diffusive decay of a pattern of surface
currents induced by the pulse.  The power law amplitude reflects
the surface geometry of the target, a property that may prove
useful in buried target classification in geophysical remote
sensing applications.

\end{abstract}

\maketitle

Remote detection and classification of buried targets is a key
goal in a number of important environmental geophysical
applications, such as toxic waste drum, and landmine and
unexploded ordnance (UXO) remediation \cite{serdp}. A common tool
used for detection of highly conducting metallic targets is the
time-domain electromagnetic (TDEM) method, in which an inductive
coil is used to transmit EM pulses into the ground.  Following
each pulse, the voltage $V(t)$ induced by the scattered field is
detected by a receiver coil \cite{GMR}. The magnitude and lifetime
of the currents induced in the target, and hence $V(t)$, increase
with its size and conductivity. Standard TDEM sensors are capable
of resolving anomalies from very small (of order 1 gram) metal
targets \cite{NH01}, and are therefore well suited to detection of
relatively large buried conducting bodies such as UXO.  However,
since TDEM is a very low frequency (typically of order 100 Hz)
method, its spatial resolution (limited by the target depth and
the sensor diameter) is also very low \cite{radar}. Therefore, the
raw signal amplitude and lifetime provide gross measures of the
target size and conductivity, but give no direct information about
its geometry and other physical characteristics that would enable
\emph{discrimination} between, say, UXO and similarly sized
clutter.

Lacking direct target geometry signatures in TDEM data (analogous
to, e.g., optical and radar images of unoccluded targets), one
seeks \emph{indirect} measures via more detailed analysis of
$V(t)$. This signal contains information about both
\emph{intrinsic} (target size, shape, geometry, and other physical
characteristics) and \emph{extrinsic} (relative target-sensor
orientation, transmitter and receiver coil geometries, pulse
waveform, etc.) properties, and the key to discrimination is the
extraction of the former from the ``background'' of the latter. We
shall show that such an analysis divides naturally into early,
intermediate and late time domains. Intermediate time is
characterized by a finite superposition of exponential decays, the
slowest of which eventually dominates and defines late time. Early
time is characterized by an essentially infinite number of
exponential decays which superimpose to generate a $1/\sqrt{t}$
universal power law divergence in $V(t)$ \cite{NMR}.  The
importance of this latter interval is greatly enhanced for ferrous
targets whose response is so slow that it may in fact comprise the
\emph{full measured range} of $V(t)$.

To focus the discussion consider the following model. At low
frequencies the dielectric function in the ground and in the
target is dominated by the its imaginary part, $\epsilon = 4\pi i
\sigma/\omega$ \cite{insulator}, where $\sigma({\bf x})$ is the dc
conductivity (in Gaussian units), and the Maxwell equations may be
reduced to a single equation for the vector potential,
\begin{equation}
\nabla \times \left(\frac{1}{\mu}
\nabla \times {\bf A} \right) +
\frac{4\pi \sigma}{c^2} \partial_t {\bf A}
= \frac{4\pi}{c} {\bf j}_S,
\label{1}
\end{equation}
with magnetic induction ${\bf B} = \nabla \times {\bf A}$ and
gauge chosen so that the electric field is ${\bf E} = -(1/c)
\partial_t {\bf A}$.  The transmitter loop is modelled by the
source current density ${\bf j}_S({\bf x},t) = I_0(t) {\bf
C}_T({\bf x})$, where the current $I_0$ consists of a periodic
sequence of rapidly terminated pulses, and ${\bf C}_T({\bf x})$
defines the transmitter loop. The magnetic field is ${\bf H} =
{\bf B}/\mu$, where $\mu({\bf x})$ is the (relative) permeability.
The conductivity and permeability are separated into background
($\sigma_b({\bf x})$, $\mu_b({\bf x})$) and conducting target
($\sigma_c({\bf x})$, $\mu_c({\bf x})$) components, where
$\sigma_c$, $\mu_c$ vanish outside the target volume $V_c$, and it
is assumed only that $\sigma_b/\sigma_c \ll 1$.

Equation (\ref{1}) is a vector diffusion equation with diffusion
constant $D = c^2/4\pi \mu \sigma$.  Typical values are $D_b = 8.0
\times 10^{10}$ m$^2$/s for a nonmagnetic background with
resistivity of 10 $\Omega$m; $D_c = 2.3 \times 10^2$ m$^2$/s for
an aluminum target with resistivity $2.8 \times 10^{-8}\ \Omega$m;
and $D_c = 4.0$ m$^2$/s for a steel target with relative
permeability 200 and resistivity $8.9 \times 10^{-8}\ \Omega$m. EM
signal propagation distance $d$ in time $t$ may be estimated via
$d \sim \sqrt{Dt}$. Early-time results will require that $\tau_b =
R^2/D_b \ll \tau_c = L_c^2/D_c$, where $R$ is the distance between
the sensor and the target, and $L_c$ is the latter's linear size:
target--sensor propagation time should be instantaneous on the
time scale of the electrodynamics of the target itself
\cite{insulator2}.  The associated condition $R \ll L_c
\sqrt{D_b/D_c}$ is easily satisfied for centimeter scale targets
at tens of meters depth, and is even less stringent for larger
targets.  The off-ramp time, $\tau_r$, for the transmitted pulse
is assumed to satisfy $\tau_r \ll \tau_c$, so that pulse
termination also occurs essentially instantaneously on the scale
of the target dynamics (no particular relation between $\tau_r$
and $\tau_b$ is required).

In order to further elucidate the various time scales in the
problem, consider the homogeneous version of (\ref{1}), valid
between pulses.  The general solution takes the form of a
superposition of exponentially decaying eigenmodes
\begin{equation}
{\bf A}({\bf x},t) = \sum_{n=1}^\infty
A_n {\bf a}^{(n)}({\bf x}) e^{-\lambda_n (t-t_0)}
\label{2}
\end{equation}
in which $t_0$ marks the beginning of the free decay window, and
the mode shapes ${\bf a}^{(n)}$ and decay rates $\lambda_n$
satisfy the eigenvalue equation
\begin{equation}
\nabla \times \left(\frac{1}{\mu}
\nabla \times {\bf a}^{(n)} \right)
= \frac{4\pi \sigma \lambda_n}{c^2} {\bf a}^{(n)}.
\label{3}
\end{equation}
These modes (which may be orthonormalized by noting that
$\sqrt{\sigma} {\bf a}^{(n)}$ are eigenfunctions of a self adjoint
operator) correspond to special current density patterns ${\bf
j}^{(n)}({\bf x}) = (\lambda_n/c) \sigma({\bf x}) {\bf
a}^{(n)}({\bf x})$ with decaying amplitude, but time-independent
spatial structure \cite{continuum}.  The spectrum $\{\lambda_n \}$
is bounded below, with fundamental decay rate $\lambda_1 \sim
1/\tau_c$ governed by the target size, but unbounded from above
\cite{MFT,W01,WL03}, with more rapidly decaying modes having
spatial structure on ever smaller scales. Since $\sigma_b \ll
\sigma_c$ currents in the background are negligible compared to
those in the target, and it follows that $\lambda_n$, as well as
the \emph{internal} structure of ${\bf a}^{(n)}({\bf x})$, ${\bf
x} \in V_c$, are essentially independent of the background
\cite{feedback}. In this sense $\lambda_n$ and ${\bf a}^{(n)}$ are
\emph{intrinsic} properties of the target. Explicitly, one finds
\begin{eqnarray}
A_n &=& \frac{4\pi}{c^2} I_n \int_{C_T}
{\bf a}^{(n)*}({\bf x}) \cdot d{\bf l}
\nonumber \\
I_n &=& \int_{-\infty}^{t_0} dt' I_0(t')
e^{-\lambda_n (t_0 - t')},
\label{4}
\end{eqnarray}
where the transmitter coil here is an idealized 1D curve $C_T$. If
the coil has $N_T$ windings then $I_0(t) = N_T i_0(T)$, where
$i_0$ is the actual current. The voltage measured in the receiver
loop is then
\begin{eqnarray}
V(t) &=& \sum_{n=1}^\infty V_n e^{-\lambda_n (t - t_0)}
\nonumber \\
V_n &=& \frac{\lambda_n N_R}{c} A_n
\int_{C_R} {\bf a}^{(n)}({\bf x}) \cdot d{\bf l},
\label{5}
\end{eqnarray}
in which $C_R$ is the idealized 1D receiver loop and $N_R$ is the
number of windings.

The excitation coefficients $A_n$ and $V_n$ depend on both
intrinsic (eigenmode) and extrinsic (transmitter/receiver loop
geometry, position, orientation, etc.) information.  Given only
$V(t)$, absent any information regarding the measurement geometry,
target classification relies entirely on the extractable subset of
decay rates $\lambda_n$.  The mathematical problem is equivalent
to the famous ``Can you hear the shape of a drum?'' (i.e., to what
extent is the shape of a struck drumhead determined by its
frequency spectrum?), but is practically much more difficult
because no analogue of the Fourier transform exists for directly
estimating the $\lambda_n$. In contrast, if detailed measurement
information is available, direct prediction of the amplitudes, and
hence of the full signal $V(t)$ is possible. A classification
scheme may then be developed based on a search for the target
model that directly minimizes the difference between the measured
and predicted data \cite{WL03}, thus circumventing the (generally
unstable) problem of direct estimation of $\{\lambda_n \}$ from
noisy data.

The number of substantially excited modes in (\ref{2}) depends on
$\tau_r$ (50--100 $\mu$s in many commercial systems). Roughly, the
terminating pulse will excite a subset (depending on the extrinsic
parameters) of those modes with $\lambda_n \alt \lambda_r \equiv
1/\tau_r$. The higher order modes will decay very rapidly, but
still contribute strongly at early time $t-t_0 = {\cal
O}(\tau_r)$. Realistically, one can hope to accurately compute
only the first few hundred modes \cite{W01,WL03}. If the largest
computable decay rate $\lambda_{\rm max}$ is smaller than
$\lambda_r$, then the interval $0 \leq t-t_0 \leq 1/\lambda_{\rm
max}$ will not be accurately modelled.  For ferrous targets it is
often the case that $1/\lambda_1$ \emph{exceeds} the measurement
window and the response is \emph{entirely} early time. The
remainder of this paper is therefore concerned with the
development of a complementary theory that deals with this
interval.  By combining this theory with the mode analysis, a
comprehensive model of the entire time-domain signal emerges.

The analysis proceeds in three steps.  (1) An ``initial
condition'' for the free dynamics, consisting of a pattern of
currents confined to the surface of the target, is computed.  (2)
The time-development of this surface current, namely its diffusion
into the interior of the target, is computed.  (3) Finally, this
solution is used to compute the external field generated at the
sensor.

\emph{Step 1:}  The rapid quenching of the transmitter current
leads to an outgoing EM pulse that scatters off the target in a
complicated way, but exits the target region by some transient
time $t_{\rm tr} = t_0 + \tau_{\rm tr}$, with $\tau_{\rm tr} =
{\cal O}(\tau_b)$.  The assumption $\tau_b \ll \tau_c$ implies
that the internal field ${\bf A}({\bf x},t_0-\tau_r)$ just prior
to the pulse termination, remains essentially fixed during the
interval $-\tau_r < t-t_0 < \tau_{\rm tr}$, responding only in a
thin shell near the boundary, $\partial V_c$.  More precisely, at
high frequencies where the target skin depth $\delta_c = \sqrt{2
D_c/\omega}$ is much smaller than the scale of tangential
variation of ${\bf A}$, the internal field near the surface, with
local normal ${\bf \hat n}$, takes the form \cite{Jackson}
\begin{equation}
{\bf A}({\bf x},\omega) = {\bf A}^\parallel({\bf r},\omega)
e^{-|z| \sqrt{-i\omega/D_c}},
\label{6}
\end{equation}
in which $z$ the coordinate along ${\bf \hat n}$, and ${\bf r}$ is
orthogonal to it, and ${\bf \hat n} \cdot {\bf A}^\parallel = 0$.
Continuity of ${\bf \hat n} \times {\bf A}$ implies that ${\bf
A}^\parallel$ is also the tangential component of the external
field.  In the time domain, (\ref{6}) becomes
\begin{eqnarray}
{\bf A}({\bf x},t) &=& {\bf A}_0({\bf x},t)
+ \Delta {\bf A}({\bf x},t)
\label{7} \\
\Delta {\bf A}({\bf x},t) &=&
\int_{t_0-\tau_r}^t dt' {\bf A}^\parallel({\bf r},t')
\frac{|z|}{t-t'} \frac{e^{-z^2/4D_c (t-t')}}
{\sqrt{4\pi D_c (t-t')}},
\nonumber
\end{eqnarray}
valid for $t-t_0 \ll \tau_c$, demonstrating the diffusion of the
signal inwards from the surface. The current density is given by
the same expression, but with ${\bf A}^\parallel$ replaced by
$\sigma_c {\bf E}^\parallel \equiv -(\sigma_c/c) \partial_t {\bf
A}^\parallel$. Integrating over $z$, at $t_{\rm tr}$ there is an
effective surface current,
\begin{equation}
{\bf K}({\bf r},t_{\rm tr}) = \int_{t_0-\tau_r}^{t_0+\tau_{\rm tr}}
\frac{dt'}{2\pi} \sqrt{\frac{\sigma_c}{\mu_c (t-t')}}
\partial'_t {\bf A}^\parallel({\bf r},t')
\label{8}
\end{equation}
confined to a thin shell with width $\sqrt{D_c (\tau_r + \tau_{\rm
tr})}$.

Equation (\ref{8}) provides a rigorous foundation for ${\bf K}$,
expressing it in terms the external field at the boundary, but the
latter has no simple form and is generally unknown.  We now
describe an alternate procedure for its direct computation via a
self-consistency argument. At time $t_{\rm tr}$ (\ref{1}) is to be
solved with ${\bf j}_S = 0$. Since all background transients have
died out, the $\sigma \partial_t {\bf A}$ term is of relative
order $D_c R^2/D_b L_c^2$ compared to the curl term and may be
dropped. It follows that ${\bf H} = \mu_b^{-1} \nabla \times {\bf
A} = -\nabla \Phi$ is the gradient of a magnetic potential
satisfying
\begin{equation}
\nabla \cdot \left(\mu_b \nabla \Phi \right) = 0.
\label{9}
\end{equation}
The solution $\Phi_0$ to this equation must satisfy appropriate
boundary conditions on $\partial V_t$, namely ${\bf \hat n} \cdot
(\mu_b {\bf H}_b - \mu_c {\bf H}_c) = 0$ and ${\bf \hat n} \times
({\bf H}_b - {\bf H}_c) = (4\pi/c) {\bf K}$ \cite{Jackson}.  In
both cases, ${\bf H}_c = \nabla \times {\bf A}_c$ is obtained from
the initial internal field ${\bf A}({\bf x},t_0-\tau_r)$ evaluated
at the boundary. The first condition imposes a unique solution on
$\Phi$ via the Neumann boundary condition
\begin{equation}
-{\bf \hat n} \cdot \nabla \Phi_0
= \frac{\mu_c}{\mu_b} {\bf \hat n}
\cdot {\bf H}_c({\bf r},z=0^-),
\label{10}
\end{equation}
with formal solution
\begin{equation}
\Phi_0({\bf x}) = \int_{\partial V_c} d^2r' g_N({\bf x},{\bf r}')
\frac{\mu_c}{\mu_b} {\bf \hat n} \cdot {\bf H}_c({\bf r}'),
\label{11}
\end{equation}
where $g_N$ is the Neumann green function satisfying $-\nabla
\cdot (\mu_b \nabla g_N) = \delta({\bf x}-{\bf x}')$ with boundary
condition ${\bf \hat n} \cdot \nabla g_N = 0$.  The second condition
determines ${\bf K}$:
\begin{equation}
{\bf K} = -\frac{c}{4\pi} {\bf \hat n}
\times (\nabla \Phi_0 + {\bf H}_c).
\label{12}
\end{equation}

\emph{Step 2:}  In order to investigate the subsequent evolution
of the surface current ${\bf K}$ we take advantage of the rapid
variation of the fields near the surface with $z$.  Thus, the
$z$-derivatives dominate (\ref{1}), and to leading order in the
small parameter $\epsilon = \sqrt{D_c (t - t_{\rm tr})/L_c^2}$ one
need only solve the one-dimensional diffusion equation
\begin{equation}
D_c \partial_z^2 {\bf E}^\perp + \partial_t {\bf E} = 0,\ z < 0,
\label{13}
\end{equation}
with initial condition ${\bf E}(t_{\rm tr}) = \sigma_c^{-1} {\bf
K} \delta(z)$.  Here ${\bf E}^\perp = {\bf E} - {\bf \hat n} ({\bf
\hat n} \cdot {\bf E})$ is the tangential part of ${\bf E}$, and
$\mu, \sigma, D$ are treated as constants on either side of the
boundary.  Since the external field varies only on the scales
$L_c,R$, to leading order one has $\partial_z {\bf E}^\perp(z=0^+)
= 0$. Continuity of ${\bf E}^\perp$ therefore imposes the Neumann
boundary condition $\partial_z {\bf E}^\perp(z=0^-) = 0$.  The
solution to (\ref{13}) is therefore
\begin{equation}
{\bf E}({\bf x},t) = {\bf E}_0({\bf x},t)
+ \frac{2}{\sigma_c} \frac{e^{-z^2/4 D_c(t-t_{\rm tr})}}
{\sqrt{4\pi D_c (t-t_{\rm tr})}},\ z < 0,
\label{14}
\end{equation}
corresponding to a diffusive Gaussian spread with rapid
$z$-dependence is on the scale $\sqrt{D_c (t-t_{\rm tr})} =
\epsilon L_c$.  By integrating with respect to time, and enforcing
the condition that ${\bf A}$ should approach the background
solution ${\bf A}_0$ for large $z/\epsilon L_c$, one obtains
\begin{eqnarray}
\Delta {\bf A}({\bf x},t) &=& \frac{4\pi
\mu_c}{c}{\bf K}({\bf r}) \left[4 D_c(t-t_{\rm tr})
\frac{e^{-z^2/4D_c (t-t_{\rm tr})}}
{\sqrt{4\pi D_c (t-t_{\rm tr})}} \right.
\nonumber \\
&&\ \ \ \ -\ \left. |z| \mbox{erfc}\left(\frac{|z|}
{\sqrt{4 D_c(t-t_{\rm tr})}} \right)\right],
\label{15}
\end{eqnarray}
where ${\rm erfc}(z) = (2/\sqrt{\pi}) \int_z^\infty e^{-s^2} ds$
is the complementary error function.  Since ${\bf K}$ is, in fact,
spread over a width $\sqrt{D_c \tau_{\rm tr}}$, equations
(\ref{14}) and (\ref{15}) are accurate only in the range
$\tau_{\rm tr} \ll t-t_0 \ll \tau_c$ where the precise microscopic
structure (\ref{7}) has been washed out by the diffusion kernel.

\emph{Step 3:}  Equation (\ref{14}) evaluated at $z=0^-$ provides
the necessary boundary condition for evaluation of the external
field to leading order in $\epsilon$.  Note that ${\bf E}(z = 0^-)
\approx {\bf K} \sqrt{4 \mu_c/\sigma_c c^2 (t - t_{\rm tr})}$
\emph{diverges,} and continuity of ${\bf E}^\perp$ leads one to
expect a corresponding divergence in the external electric field.
We exhibit this formally through a correction $\Delta \Phi$ to the
magnetic potential. Thus, the normal component of the curl of
(\ref{15}) leads to the boundary value ${\bf \hat n} \cdot {\bf
B}(z=0^-) = {\bf \hat n} \cdot ({\bf B}_0 + \Delta {\bf B})$,
where
\begin{equation}
{\bf \hat n} \cdot \Delta {\bf  B} = 4 \sqrt{t-t_{\rm tr}}
{\bf \hat n} \cdot \nabla \times
\left(\sqrt{\mu_c/\sigma_c} {\bf K} \right).
\label{16}
\end{equation}
involves only derivatives with respect to the tangential
coordinate ${\bf r}$, and is valid even if $\mu_c, \sigma_c$ vary
on the scale $L_c$.  We therefore obtain $\Phi = \Phi_0 + \Delta
\Phi$, with boundary condition $-\mu_b {\bf \hat n} \cdot \Delta
\Phi = {\bf \hat n} \cdot \Delta {\bf B}$, and hence to ${\cal
O}(\epsilon)$,
\begin{equation}
\Delta \Phi({\bf x}) = \int d^2r' g_N({\bf x},{\bf r}')
\frac{1}{\mu_b} {\bf \hat n} \cdot \Delta {\bf B}({\bf r}'),
\label{17}
\end{equation}
which is proportional to $\sqrt{t-t_{\rm tr}}$.  The correction to
the external vector potential is obtained by solving the auxiliary
pair of equations
\begin{eqnarray}
\nabla \times \Delta {\bf A} &=& -\mu_b \nabla \Delta \Phi
\nonumber \\
\nabla \cdot (\sigma_b \Delta {\bf A}) &=& 0,
\label{18}
\end{eqnarray}
and is clearly also proportional to $\sqrt{t-t_{\rm tr}}$.  The
electric field correction $\Delta {\bf E} = -(1/c)
\partial_t \Delta {\bf A} \propto (t-t_{\rm tr})^{-1/2}$ therefore
has the promised square root early time divergence. Measurements
of magnetic field or voltage (via the time derivative of the of
the integral of the magnetic flux through receiver loop area)
follow directly from (\ref{17}).

We end by illustrating the early time behavior using exact
analytical results for a homogeneous sphere of radius $a$ in a
homogeneous background \cite{W03}.  We consider also an initial
static transmitted field, so that the initial magnetic field is
everywhere described by a scalar potential.  The initial solution
is a superposition of spherical harmonics,
\begin{equation}
\Phi^{lm}_{\rm init} = Y_{lm}
\left\{\begin{array}{ll}
(r/a)^l, & r < a \\
b_{\rm init}^{lm} (r/a)^l
+ c_{\rm init}^{lm} (a/r)^{l+1}, & r > a.
\end{array} \right.
\label{19}
\end{equation}
with $c_{\rm init}^{lm} = 1 - b_{\rm init}^{lm} = (1-\mu_c/\mu_b)
\frac{l}{2l+1}$ ($l=1$ corresponds to the standard case of a
uniform illumination field, leading to a dipole response). At
$t_{\rm tr}$ the $r < a$ solution remains the same, while
$b_0^{lm}$ vanishes. The boundary condition (\ref{10}) then leads
to $\Phi_0^{lm} = c_0^{lm} Y_{lm}$, with $c_0^{lm} =
-l\mu_c/(l+1)\mu_b$.  From (\ref{12}, surface current is,
\begin{equation}
{\bf K}^{lm} = \frac{ic}{4\pi a}
\left(1+\frac{l}{l+1} \frac{\mu_c}{\mu_b} \right)
\sqrt{l(l+1)} {\bf X}_{lm}.
\label{20}
\end{equation}
From (\ref{16}) and (\ref{17}) one then obtains,
\begin{eqnarray}
\Delta \Phi^{lm} &=& -\phi_l(t) \left(\frac{a}{r}
\right)^{l+1} Y_{lm}
\label{21} \\
\phi_l(t) &\equiv& \frac{\mu_c l}{\mu_b a}
\left(1+\frac{l}{l+1} \frac{\mu_c}{\mu_b} \right)
\sqrt{\frac{4D_c(t-t_{\rm r})}{\pi}},
\nonumber
\end{eqnarray}
and the external fields are given by
\begin{equation}
\Delta {\bf B}^{lm} = i \mu_b \sqrt{\frac{l+1}{l}} \phi_l(t)
\nabla \times \left[\left(\frac{a}{r} \right)^{l+1}
{\bf X}_{lm}\right]
\label{22}
\end{equation}
\begin{equation}
\Delta {\bf A}^{lm} = i \mu_b \sqrt{\frac{l+1}{l}} \phi_l(t)
\left(\frac{a}{r} \right)^{l+1} {\bf X}_{lm},
\label{23}
\end{equation}
which each display $\sqrt{t}$ cusps, while
\begin{equation}
\Delta {\bf E}^{lm} = -\frac{i \mu_b}{c} \sqrt{\frac{l+1}{l}}
\frac{\phi_l(t)}{2(t-t_{\rm tr})} \left(\frac{a}{r} \right)^{l+1}
{\bf X}_{lm},
\label{24}
\end{equation}
has the $1/\sqrt{t}$ divergence. Here ${\bf X}_{lm} =
-i[l(l+l)]^{-1/2} {\bf x} \times \nabla Y_{lm}$ are the vector
spherical harmonics \cite{Jackson}.  The spatial decay rate of the
signal increases with $l$, but all harmonics have the same
universal power law time-dependence.

The author is indebted to E. M. Lavely for numerous discussions.
The support of SERDP, through contract No.\ DACA 72-02-C-0029, is
gratefully acknowledged.


\begin{thebibliography}{9}
\bibitem{radar} Low frequency is crucial for reasonable ($> 10$ m)
depth sensitivity.  At ground penetrating radar frequencies, where
the short wavelength ($< 10$ cm) ensures good spatial resolution,
the EM skin depth is at best a few meters, and much smaller still
in wet soil.  In addition, the dielectric contrast between the
ground and the target (and similarly sized clutter, such as rocks)
is greatly reduced, thus also reducing target classification
capability.
\bibitem{insulator} This actually need only be true inside the
target.  The background may be insulating ($\epsilon = \epsilon'$
real), with no significant change in the results.  It is required
only that signals, be they diffusive or wavelike, propagate much
faster in the background than in the target.
\bibitem{insulator2} In an insulating background one has instead
$d \sim c t/\sqrt{\epsilon_b'}$, and the much less stringent
condition $R/c \ll \tau_c$.
\bibitem{feedback} The external field, however, will be distorted
by background variations, but there is negligible feedback on the
internal field. This may be formalized via the Green function
integral formulation of (\ref{3}) which reduces the eigenvalue
equation to one for the internal field alone.
\end{thebibliography}
\end{document}